\documentclass[manuscript,screen]{acmart}
\usepackage{balance}
\copyrightyear{2021}
\acmYear{2021}
\setcopyright{acmlicensed}\acmConference[PEARC '21]{Practice and Experience in Advanced Research Computing}{July 18--22, 2021}{Boston, MA, USA}
\acmBooktitle{Practice and Experience in Advanced Research Computing (PEARC '21), July 18--22, 2021, Boston, MA, USA}
\acmPrice{15.00}
\acmDOI{10.1145/3437359.3465578}
\acmISBN{978-1-4503-8292-2/21/07}

\AtBeginDocument{%
  \providecommand\BibTeX{{%
    \normalfont B\kern-0.5em{\scshape i\kern-0.25em b}\kern-0.8em\TeX}}}

\usepackage{hyperref}

\usepackage{siunitx}

\newcommand{\RIKEN}{RIK$\Xi$N~}

\begin{document}

\title[Ookami]{Ookami: Deployment and Initial Experiences}


\author{Andrew Burford}
\email{andrew.burford@stonybrook.edu}
\author{Alan C.\ Calder}
\email{alan.calder@stonybrook.edu}
\author{David Carlson}
\email{david.carlson@stonybrook.edu}
\author{Barbara Chapman}
\email{barbara.chapman@stonybrook.edu}
\author{Fırat Coşkun}
\email{firat.coskun@stonybrook.edu}
\author{Tony Curtis}
\email{anthony.curtis@stonybrook.edu}
\author{Catherine Feldman}
\email{catherine.feldman@stonybrook.edu}
\author{Robert J.\ Harrison}
\email{robert.harrison@stonybrook.edu}
\author{Yan Kang}
\email{yan.kang@stonybrook.edu}
\author{Benjamin Michalowicz}
\email{benjamin.michalowicz@stonybrook.edu}
\author{Eric Raut}
\email{eric.raut@stonybrook.edu}
\author{Eva Siegmann}
\email{eva.siegmann@stonybrook.edu}
\author{Daniel G.\ Wood}
\email{daniel.g.wood@stonybrook.edu}
\affiliation{%
\institution{Institute for Advanced Computational Science, Stony Brook University}
\streetaddress{1 Nicholls Rd.}
\city{Stony Brook}
\state{NY}
\country{USA}}

\author{Robert L.\ DeLeon}
\email{rldeleon@buffalo.edu}
\author{Mathew Jones}
\email{jonesm@buffalo.edu}
\author{Nikolay A.\ Simakov}
\email{nikolays@buffalo.edu}
\author{Joseph P.\ White}
\email{jpwhite4@buffalo.edu}
\affiliation{%
  \institution{Center for Computational Research, University at Buffalo}
  \streetaddress{701 Ellicott St.}
  \city{Buffalo}
  \state{NY}
  \country{USA}
}

\author{Dossay Oryspayev}
\email{doryspaye@bnl.gov}
\affiliation{%
  \institution{Brookhaven National Laboratory}
  \streetaddress{PO Box 5000}
  \city{Upton}
  \state{NY}
  \country{USA}
}

\renewcommand{\shortauthors}{Burford, et al.}

\begin{abstract}
  Ookami~\cite{ookamiurl} is a computer technology testbed supported by the United States National Science Foundation. It provides researchers with access to the A64FX processor developed by Fujitsu~\cite{fujitsuA64FXurl} in collaboration with \RIKEN~\cite{rikenurl,rikenccsurl} for the Japanese path to exascale computing, as deployed in Fugaku~\cite{fugakuurl}, the fastest computer in the world \cite{fugakufastest}. By focusing on crucial architectural details, the ARM-based, multi-core, 512-bit SIMD-vector processor with ultrahigh-bandwidth memory promises to retain familiar and successful programming models while achieving very high performance for a wide range of applications. We review relevant technology and system details, and the main body of the paper focuses on initial experiences with the hardware and software ecosystem for micro-benchmarks, mini-apps, and full applications, and starts to answer questions about where such technologies fit into the NSF ecosystem.
\end{abstract}



\maketitle

\section{Introduction}

 Fugaku \cite{fugakuurl}, the presently fastest computer in the world, and its associated processor technology \cite{fujitsuA64FXurl} have achieved an {\em unprecedented sweep} of five major benchmarks \cite{fugakufastest} --- Top 500 \cite{fugakutop500}, Graph-500 \cite{fugakugraph}, HPCG \cite {hpcgurl}, HPL-AI \cite{hplaiurl}, and, in 2019, Green-500 \cite{fugakugreen}. The Fujitsu Tofu-D communication fabric \cite{fujistutofud} certainly contributes to these successes.  But central is the novel A64FX processor \cite{fujitsuA64FXurl}, which is a significant reversal in the global trend to supercomputers being powered by graphical processing units (GPUs) that typically burden the programmer with proprietary programming models and complex memory hierarchies. The Ookami test bed gives US researchers free access to evaluate this exciting processor technology.  It is also an opportunity to explore the potential of this technology for deployment in multiple settings including at extreme scale in national supercomputer centers or on campus at the network edge. 
 
 The 48-core, 64-bit ARM processor is the first to deploy the Scalable Vector Extension (SVE) SIMD-vector instruction set, employing 512-bit wide vectors matched with 32~Gbyte of high-bandwidth memory (1~Tbyte/s). Designed specifically for leadership supercomputers, this processor+memory system promises to retain familiar and successful programming models while achieving very high performance for a wide range of applications. It supports 64/32/16-bit floating-point representations and fast partial dot-product of 8-bit integers to 32-bit results, and hence enables both HPC and big data. As demonstrated by the HPCG and Top500 benchmarks, A64FX offers strong performance on memory-intensive applications such as sparse-matrix solvers, while also being competitive on floating-point-intensive codes. We will see below that for some software this transformational performance is available nearly out of the box --- MPI+OpenMP vectorized code should just compile and immediately run well, with additional performance possible from tuning. 

The applications that will benefit most are those sensitive to memory bandwidth with a memory footprint less than 32 GB/node --- these two features capture the vast majority of NSF’s workload. In 2017, team members conducted a workload study \cite{simakovworkload} of NSF’s computing resources including large-memory queues.  The study found that 86\% of all jobs (corresponding to 85\% of cycles) will fit within the available memory per node (including O/S) and that the majority of jobs are memory-bandwidth intensive. 


\section{Ookami technology and system architecture}

\subsection{System configuration}

The Ookami HPE Apollo 80 system has 174 1.8~Ghz A64FX compute nodes each with 32GB of high-bandwidth memory and a 512 Gbyte SSD. The latter is a post-market addition made by the project. A high-performance Lustre filesystem employing the Cray ClusterStor E1000 provides about 0.8Pbyte storage.  Nodes are connected by an HDR 200~Gbit/s Infiniband network configured as a full fat tree. Using a Y-cable, the HDR 100 Gbit/s NICs of two compute nodes are connected to each HDR 200 Gbit/s switch port. This recognizes the bandwidth limitation of the on-chip PCIe interface and also reduces cable density. The login nodes were originally x86 but are now dual socket ThunderX2 with 256~Gbyte memory to eliminate most cross-compilation issues. In addition, Ookami includes servers powered by dual-socket AMD Rome, Intel SkyLake, and Intel Haswell (two with NVIDIA V100 GPUs) for performance comparisons across architectures.

\subsection{ARM background}

ARM is a RISC processor designed by ARM Ltd. but specialized and manufactured by other companies leading to much greater diversity than exists for X86. Annual processor production of ARM is well over 20 billion, which dwarfs X86 at 3-400 million \cite{satoA64FX}. 
The development of 64-bit and other features with a deliberate push, including by the U.S. DOE FastForward program, made the processor relevant to HPC. 
More recently, systems on the Top500 \cite{Ast} have been powered by the 64-bit Cavium ThunderX2 specifically optimized for HPC including 128-bit SIMD units and more cores. A detailed evaluation \cite{McS} of a Cray XC50 system (dual-socket ThunderX2), examined a wide range of applications relevant to NSF users with four compelling conclusions:
(1) ARM can now provide performance competitive with mainstream processors; (2) with significant increases in price performance; (3) most benchmarks ran successfully out of the box; and (4) with no special tuning necessary for high performance. 
This closely mirrors our experience with A64FX, with the exception that only well-vectorized applications attain a substantial fraction of peak speed on A64FX primarily due to (1) the 512-bit vector width that is 4x that of NEON, and (2) the relatively higher latency of many instructions.

\begin{table*}[t]
  \caption{Available compilers.}
  \centering
  \begin{tabular}{l c c c}
    \hline \hline
    Compiler
    & Compiler names & Version  & Vectorization Flag \\
    \hline
    GCC & gcc / g++ / gfortran & 10.2.1 &  -mcpu=A64FX  \\ 
    Arm & armclang / armclang++ / armflang & 20.3 & -mcpu=A64FX -armpl=sve  \\
    LLVM & clang / clang++ / flang & 11.0.0 & -mccpu=A64FX  \\
    Cray & cc / CC / ftn & 10.0.1 &  -h vec3 \\
    NVIDIA & nvc / nvc++ / nvfortran & 21.2.0 &  -Mvect[=options]  \\
    Fujitsu (will be installed soon) & fcc / FCC / frt & 4.5.0 &  -KSVE  \\
    \hline
  \end{tabular}
  \label{tab:compilers}
\end{table*}

\subsection{Processor architecture}

The A64FX processor comprises 4 non-uniform memory access (NUMA) regions or core memory groups (CMGs) with on-package, stacked, high-bandwidth memory (8 Gbyte). A CMG consists of 12 cores\footnote{The processor in Fugaku includes an additional core for O/S services.} with private L1 cache (64~Kbyte), a shared L2 cache (8~Mbyte), and a memory controller. The four CMGs keep cache coherency by ccNUMA with an on-chip directory. A cross-bar connection in a CMG maximizes high efficiency for throughput of the L2 cache (>115 GB/s/core read, >57 GB/s/core write). Out-of-order mechanisms in cores, caches, and memory controllers maximize the utilization of bandwidth at each level. The cache line is 256 bytes.

The A64FX core has a superscalar architecture with out-of-order execution and branch prediction. Peak execution rate is provided by issuing 2x512-bit wide SIMD FMA + predicate op + 4x ALU (shared with 2x address generation) instructions per cycle. The theoretical peak double precision speed in FLOP/s is computed as 1.8~GHz$\times$2~FMA/cycle$\times$2~FLOPs/FMA$\times$8 64-bit words/vector~=~57.6 GFLOP/s/core. Rates for other precisions are obtained by adjusting the number of words per vector. Machine learning applications are enabled by the FP16 high performance and the INT8 partial dot product that computes the elementwise product of vectors of 8-bit integers, reducing in groups of 4 into a vector of 32-bit results.

Each cycle 2x512-bit SIMD load or 1x512-bit SIMD store are supported and are enabled by the L1 data cache that can sustain 128 bytes/cycle even for unaligned SIMD load. The L1 “combined gather” doubles the gather (indirect) load’s data throughput, when target elements are within a 128-byte aligned block for a pair of two registers, similar to the coalescing of out-of-order loads within a warp on some GPGPUs. The sector cache (scratchpad) feature is a valuable optimization for a library writer by enabling the programmer to modify how the cache replacement policy treats data, including forcing data to be kept in cache. There are 6 hardware barriers per CMG.


\subsection{Scalable vector extension (SVE)}

SVE supports vector widths from 128 to 2048 bits in 128-bit increments and was designed for HPC. With full per-lane predication, it superficially appears similar to Intel's AVX/AVX2/AVX512 instruction sets, however, there are important differences.  SVE instructions can be generated to be vector-length agnostic (VLA), enabling a binary to run on any implementation regardless of vector width, and eliminating the need for loop tails. The instruction set also includes gather/scatter, vector partitioning, software-managed speculation, horizontal reductions, and scalarized intra-vector sub-loops, which all facilitate compilers vectorizing more loops.

\section{Software environment}

The operating system on Ookami is the industry-standard and stable Linux distribution CentOS 8, and all features and tools work as expected. Indeed, outside of different performance tuning concerns one usually has little awareness of being on an A64FX platform. The industry standard Bright Cluster Manager, which has supported ARM for many years, is used to provision the hardware, operating system, and workload manager from a unified interface. 
%
%
SLURM (Simple Linux Utility for Resource Management, version 19.05.7) is employed to schedule jobs and manage the resources with nodes allocated exclusively to one job. Fair-share scheduling is presently used, rather than quotas, in order to encourage utilization. There are various partitions available covering different job duration, job sizes and architectures.
%
%
Following our practice on other clusters at Stony Brook, each user has a personal home and scratch directory that are forced to be private. Each project has a shared directory accessible only by project members. The home and project directories are backed up daily.

\subsection{Development environment}

Users have access to a continuously growing software stack, organized into modules to simplify use, management, and version control.  The available compilers are listed in Table \ref{tab:compilers}. All of these, except for the NVIDIA and mainstream LLVM compilers, are capable of generating optimized SVE code. However, there are significant differences in capability and performance. We do not yet have substantial experience with the Fujitsu compiler.  The ARM and Cray compilers can generate SVE instructions for loops containing standard math functions, with the Cray compiler typically generating faster code. However, the Cray C++ compiler front end is limited in its support for modern C++ standards, and, similarly for the ARM Fortan compiler front end. Only the ARM and recent GNU compilers support the SVE intrinsics, with the GNU compiler typically generating faster code.  The ARM, Cray, and recent GNU compilers can all vectorize "if-tests" using the SVE predicated instruction set.  Starting with version 10.2.1, the GNU chain is aware of the A64FX processor, and similarly for the upcoming LLVM version 11.0.0 that will add support for SVE intrinsics but will still lack SVE autovectorization. 

Thus, for C and C++ the choice of compiler is complicated. Code using recent language standards or SVE intrinsics must use ARM or GNU, with the former recommended for kernels using math functions and the latter for SVE intrinsics. For FORTRAN, the Cray compiler is typically the best choice for both performance and standard compliance.  

Other available modules include anaconda, archiconda, armpl, libsci, lapack, openblas, fftw, papi, cmake, etc. Various MPI versions are available, e.g. openmpi 4.0.5 and 4.1.0 and openshemem (Section \ref{sec:OpenSHMEM}).

\subsection{Performance analysis and debugging}

Multiple toolchains enable the user to analyze and optimize their code. Arm Forge includes the Arm ddt debugger, the profiling tool ARM MAP and ARM performance reports for advanced reporting capabilities. The Cray toolchain includes CrayPat, Apprentice and Reveal tools for performance analysis/tuning, and the debugging tools gdb4hpc (general debugger), valgrind4hpc (for detecting memory leaks and errors in parallel applications) and atp (first-line tool to diagnose crashing applications). The GNU gdb command-line debugger is also available. The Fujitsu toolchain is not yet installed on Ookami, but will be provided in the near future.

\section{Initial experience}

Our standing joke is that the system is "ARM-less" (i.e., "harmless") in that standard-compliant applications in FORTRAN, C, or C++ simply compile and run out of the box, once mundane issues such as compiler flags and library paths have been addressed.  This is due to (1) the standard and complete Linux distribution, (2) the extensive selection of standard-compliant tool chains, (3) a growing library of linear algebra and scientific kernels, and (4) the availability of multiple MPI implementations (Cray, MVAPICH, OpenMPI) all optimized for A64FX and SVE.  More nuanced is what must be done to obtain high performance from the processor along with selection of the appropriate tool chain(s), as discussed above.
Early concerns included that InfiniBand performance would be inferior due to the depth of instruction pipelines and cache architecture --- these have proven unfounded (Section \ref{sec:OpenSHMEM}).

Performance measurements deliver highly reproducible results between runs on the same or different sets of nodes.  For instance, whole-node DGEMM benchmarks run by HPE/Cray during the burn-in phase detected two slow nodes, which were replaced, and the performance of all other nodes had a spread reproducible less than 1\%.  This is due to the fixed clock speed and absence of turbo modes or thermal throttling.  Multi-node runs are similarly reproducible due to fixed clock speed and the full fat tree that eliminates network contention.  

The most important early discovery was the need to install an SSD on all nodes to store the root file system.  As shipped, only two nodes targeted for software development and debugging had SSDs.  All other nodes had a memory-resident system image --- but this consumed \textasciitilde 14~Gbyte instead of the anticipated \textasciitilde 2~Gbyte.  This turned out to be due to the 64~Kbyte page size, which is the minimum space consumed by even a 1~byte file. The SSDs open the door to new data-intensive applications, and in the near future may be upgraded from 512~Gbyte to 2~Tbyte.

So far the system has been very reliable.  The two slow nodes were were replaced along with one NIC and 3 SSDs. Unscheduled outages have been due to power failures and non-A64FX-specific system software including a memory leak in a SLURM demon.

\begin{table}[b] 
  \caption{OSU MPI Benchmarks}
  \centering
  \begin{tabular}{lllll}
  \hline \hline
     Benchmark & Msg Size & Est. & Measured \\
     \hline
Two-sided uni-dir B/W &	1MB & 12.3 GB/s                & 12.3      \\
Two-sided bi-dir B/W & 	1MB & 24.4 GB/s                & 19.4      \\
Two-sided latency &     8B  &  \SI{1.8}{\micro\second} & 2.7       \\
Put latency &           8B  &  \SI{2.8}{\micro\second} & 3.9       \\
Put BW &               	1MB &  12 GB/s                 & 12.3       \\
Get latency &           8B  &  \SI{3.8}{\micro\second} & 5.2       \\
Get BW &               	1MB &  9.5 GB/s                & 12.3      \\
Injection Rate &       	1MB &  12.3 GB/s               & 12.3      \\
Saturation Point &     	1MB &  1 thread                & 1         \\
\hline
    \end{tabular}
    \label{tab:osu-mpi}
\end{table}

\subsection{Micro and standard benchmarks}

\subsubsection{Single-node DGEMM and STREAM}

Single-node DGEMM was measured by Cray on Ookami at 24.2~TFLOP/s, which is 88\% of peak single processor speed. We are anticipating this to increase as Cray scilib is further optimized, and when we get access to the Fujitsu library.  Fujitsu has reported 94\% of peak single processor speed.  At SBU, we have developed a code generator for AVX2, AVX512, NEON, and SVE instrinsics for the matrix-transpose times matrix kernel that is important in high-order spectral-element methods.  This code attains 53.24~GFLOP/s on a single core, which is 92.4\% of peak speed.

Single-node STREAM TRIAD was measured by Cray on Ookami at 830~Gbyte/s, and a simple compiled DAXPY kernel with careful binding of threads+memory to cores achieved 840 Gbyte/s. Manually-tuned kernels were needed to achieve close to peak L1 bandwidth.  About six threads are required to saturate the memory bandwidth of a single CMG.  The observed 840~Gbyte/s is 84\% of peak memory bandwidth, 93\% of what we understand to be the L2-memory bandwidth, and is about 5x a dual-socket Intel Skylake.  The measured single-core DAXPY bandwidth is 53 Gbyte/s, which is 2.5X Intel SkyLake. This memory bandwidth contributes greatly to the empirically observed strong thread scaling within a node.

\subsubsection{OSU Benchmarks with MPI and OpenSHMEM}  \label{sec:OpenSHMEM}
The Message Passing Interface~\cite{MPI:Forum,Dongarra1995AnIT} (MPI) is a widely-used parallel programming library.  It includes both two-sided rendezvous-based communication, and one-sided Remote Memory Access (RMA).  OpenSHMEM~\cite{chapman2010introducing,openshmem-website} is a Partitioned Global Address Space (PGAS) library that focuses on one-sided, decoupled, communication and uses RMA to expose memory directly to network interconnects such as Infiniband. The Ohio State University benchmark suite~\cite{OSU:bench} tests bandwidth and latency for both point-to-point and collective operations using MPI and OpenSHMEM (UPC/UPC++ were not tested).  We used Open-MPI~\cite{Open-MPI} (version 4.0.5) as it provides both MPI and OpenSHMEM in the same package.

Of particular interest was the point-to-point latency and bandwidth performance as we wanted to investigate the thread-to-network behavior on the A64FX processor.  Tests were between 2 nodes, 1 rank or processing element (PE) per node.   In Table~\ref{tab:osu-mpi} are the results for the MPI one- and two-sided tests. The "put" and "get" latencies are slightly higher than anticipated. In Table~\ref{tab:osu-openshmem} are the results for the OpenSHMEM tests.  Included here are representative latencies recorded for the 1-sided RMA operations, the posting and fetching atomics, and swaps.
The estimated time comes from comparisons made on a similar A64FX cluster~\cite{ORNL:Wombat} at Oak Ridge National Laboratory.


\begin{table}[b] 
  \caption{OSU OpenSHMEM Benchmarks: Latency}
  \centering
  \begin{tabular}{lllll}
  \hline \hline
       Benchmark & Est. & Measured \\
  \hline
Put (8B)                & \SI{5.4}{\micro\second} & 4.5 \\
Get (8B)                & \SI{4.2}{\micro\second} & 4.1 \\
Atomics (add)           & \SI{0.0}{\micro\second} & 0.8 \\
Fetching atomics (fadd) & \SI{5.2}{\micro\second} & 5.2  \\
Swap                    & \SI{4.4}{\micro\second} & 4.3 \\
\hline
    \end{tabular}
    \label{tab:osu-openshmem}
\end{table}


Installing the open-source Open-MPI and
MVAPICH2~\cite{mvapich2-website}
worked "out of the box".  Both are configured to use the Infiniband network and the on-node cross-memory-attach provided by
KNEM~\cite{knem-website}
and
XPMEM~\cite{xpmem-website}.
Our OpenSHMEM work involves a separate, reference, implementation of OpenSHMEM (although based on UCX~\cite{OpenUCX} like Open-MPI). This installed without issue as we have worked with ARM on the OpenSHMEM project for a number of years, and it has components in common with Open-MPI, and can take advantage of the on-node modules KNEM and XPMEM. Compilation with the stock GCC, a new GCC with more SVE support, and the ARM compilers were also without issue. UCX includes a number of ARM-specific optimizations, e.g. low-power wait states, high-precision timers, and vectorized memory copy.

\subsubsection{HPC Challenge Benchmark suite}
\begin{figure}[ht!]
    \centering
    \includegraphics[width=\columnwidth]{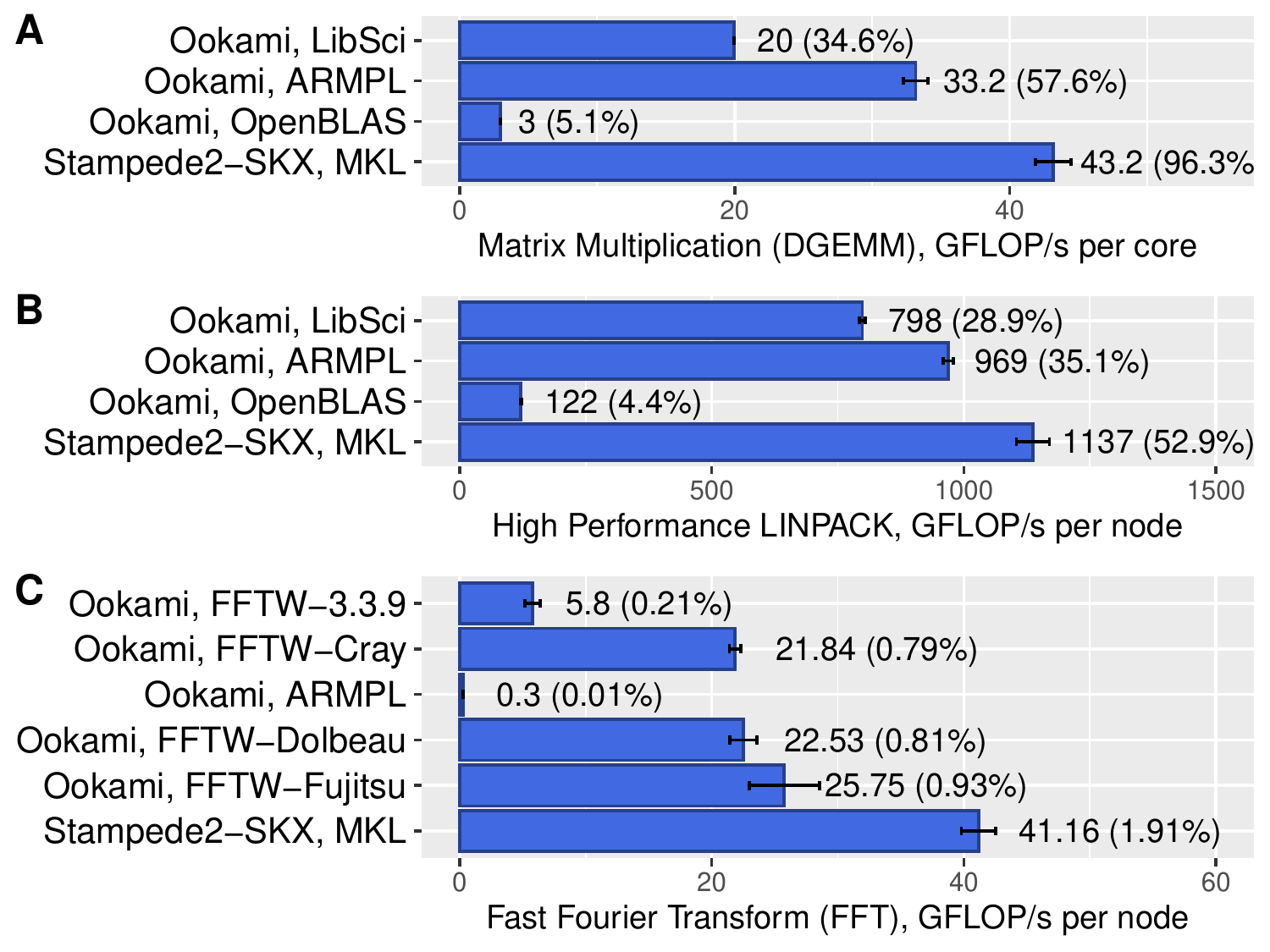}
    \caption{HPCC benchmark. Performance is shown in GFLOP/s, a number next to the bar corresponds to absolute value, and value in paranthesis corresponds to percentage of theoretical FLOP/s. A) Matrix-matrix multiplication, executed on a single core, B) High Performance LINPACK (HPL), executed on a single node, C) FFT, executed on a single node}
    \label{fig:HPCC}
\end{figure}

(HPCC)~\cite{hpcc} combines a large number of benchmarks allowing a convenient way to test multiple subsystems of HPC resource in a single run. Several of the tests rely on linear algebra and FFT libraries that are often used within scientific and engineering applications.  To benchmark these important libraries, we concentrated on matrix-matrix multiplication, High-Performance LINPACK (HPL)~\cite{hpl}, and Fast Fourier Transformation (FFT) benchmarks. Several linear algebra libraries are available on Ookami: Cray LibSci, ARM Perfomance Library (ARMPL), and OpenBLAS. The first two already have SVE support, while the last one does not. These libraries' performance is shown in Figure~\ref{fig:HPCC}.A-B. ARMPL has the best performance while OpenBLAS has the smallest due to lack of SVE support.

To HPCC we added a test of the most recent version of FFTW as well as other libraries providing the FFTW-3 API.  Within the FFT benchmark, we compared ARMPL and several version of FFTW library: official (no SVE support), provided by Dolbeau, Cray, and Fujitsu. The results are shown in Figure~\ref{fig:HPCC}.C. ARMPL (20.3) show extremely low FFT performance, suggesting falling back to a completely non-SIMD version.
The Fujitsu version of FFTW shows the best results, followed by Dolbeau and Cray. It is worth mentioning that Dolbeau developed his version of FFTW without access to actual hardware using emulators; and two other versions with SVE support are most likely based on his work.

\subsection{Mini-apps and Applications}

\subsubsection{Minimod}

Minimod \cite{meng2020minimod} is a seismic modeling mini-app developed by Total. Minimod extracts the stencil computation from a production seismic imaging application. The stencil is used to numerically solve the acoustic wave equation. The goal of Minimod is to offer a benchmark to test new and emerging hardware and programming models for geophysics applications.

\begin{figure}[ht!]
    \centering
    \includegraphics[width=\columnwidth]{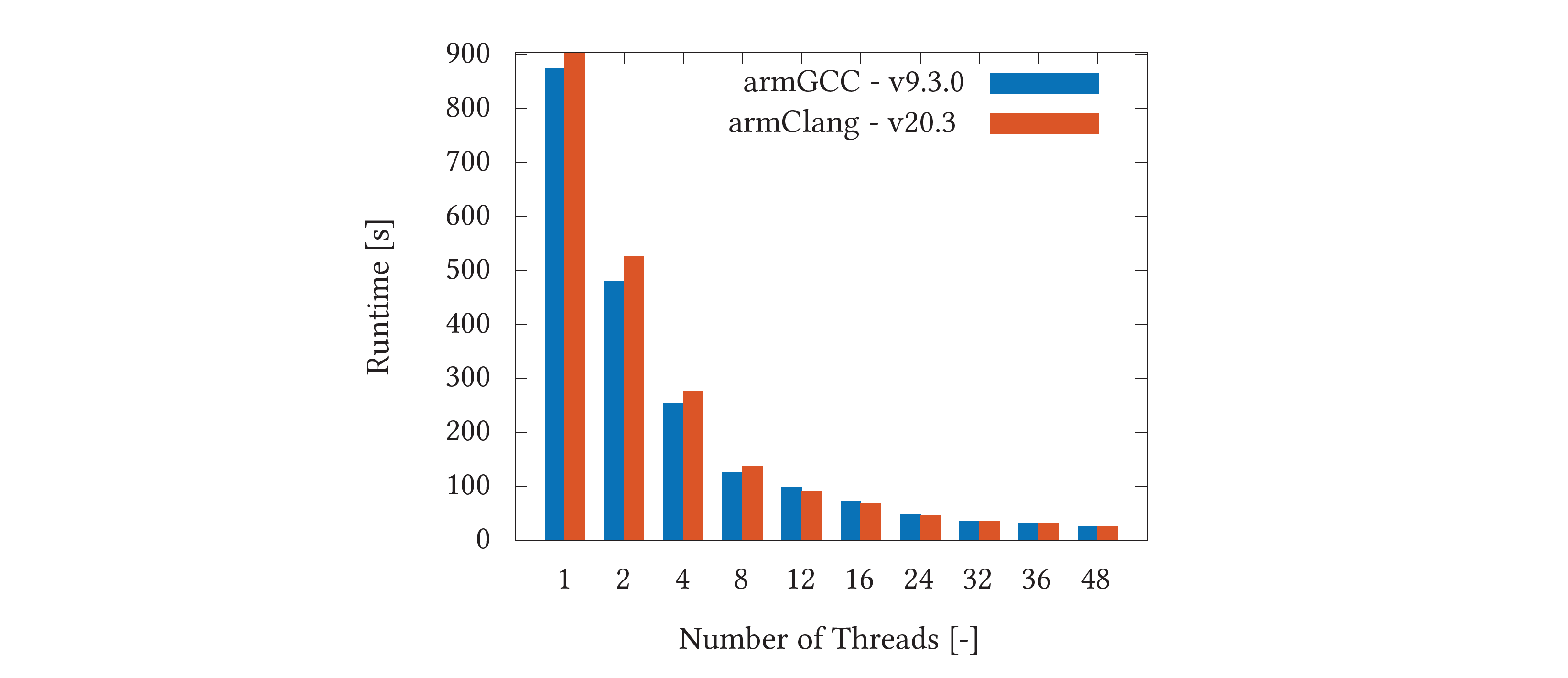}
    \caption{Minimod: Runtime results between 2 different compilers with SVE optimizations on Ookami}
    \label{fig:Minimod-tasksxy}
\end{figure}

We evaluated the Minimod OpenMP loop-based and task-based implementations \cite{EricIWOMP2020} on Ookami. Results are shown in Figure \ref{fig:Minimod-tasksxy} for the task-based implementation. While armgcc shows a performance improvement over armclang at low thread counts, at higher counts the advantage is mostly eliminated.

\begin{figure}[ht!]
    \centering
    \includegraphics[width=\columnwidth]{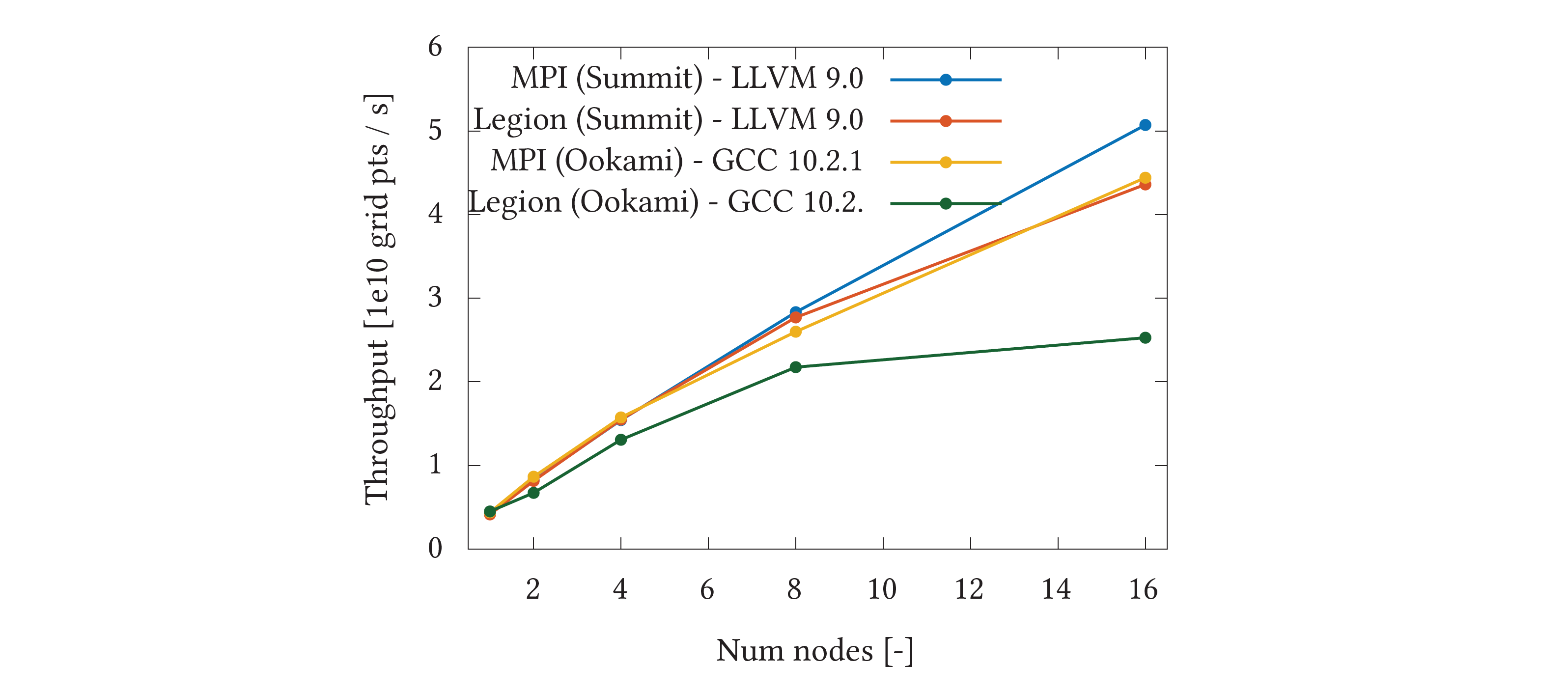}
    \caption{Minimod: Comparison of Legion and MPI throughput on Ookami and Summit}
    \label{fig:Minimod-legion}
\end{figure}

We also ported \cite{raut2021porting} Minimod to use the Legion task-based parallel programming framework \cite{bauer2012legion} and compared the result to MPI. We also compared the performance to Summit using only Summit's POWER9 CPUs. Throughput is shown in Figure \ref{fig:Minimod-legion}.
Although the MPI version of Minimod on Ookami (MVAPICH 2.3.4) achieves throughput and speedup comparable to Summit, the Legion version (Figure \ref{fig:Minimod-legion}) suffers a performance penalty in both weak scaling and strong scaling --- we are still investigating the cause.

\subsubsection{SWIM}
\begin{figure}[ht!]
    \centering
    \includegraphics[width=\columnwidth]{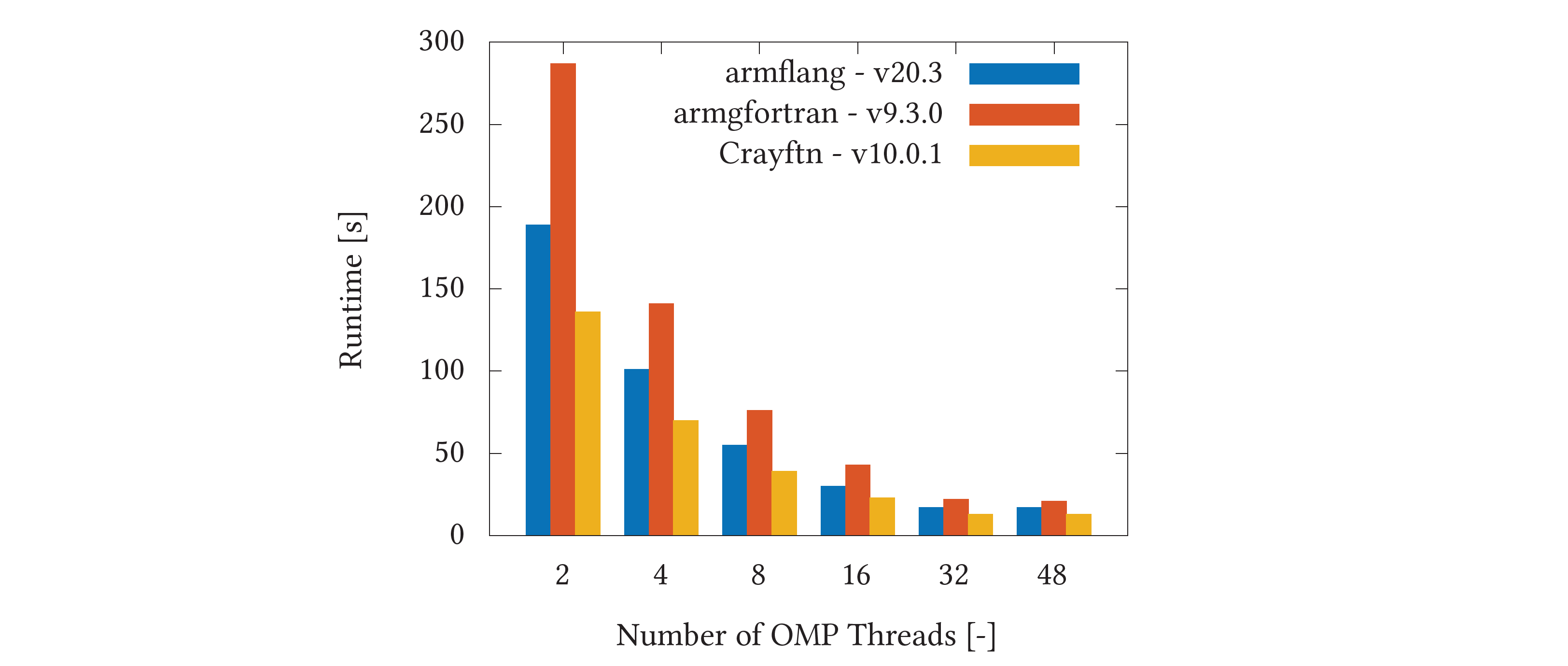}
    \caption{SWIM: Runtime results for three compilers with SVE optimizations on Ookami.}
    \label{fig:SWIM-Ookami}
\end{figure}

\begin{figure}[ht!]
    \centering
    \includegraphics[width=\columnwidth]{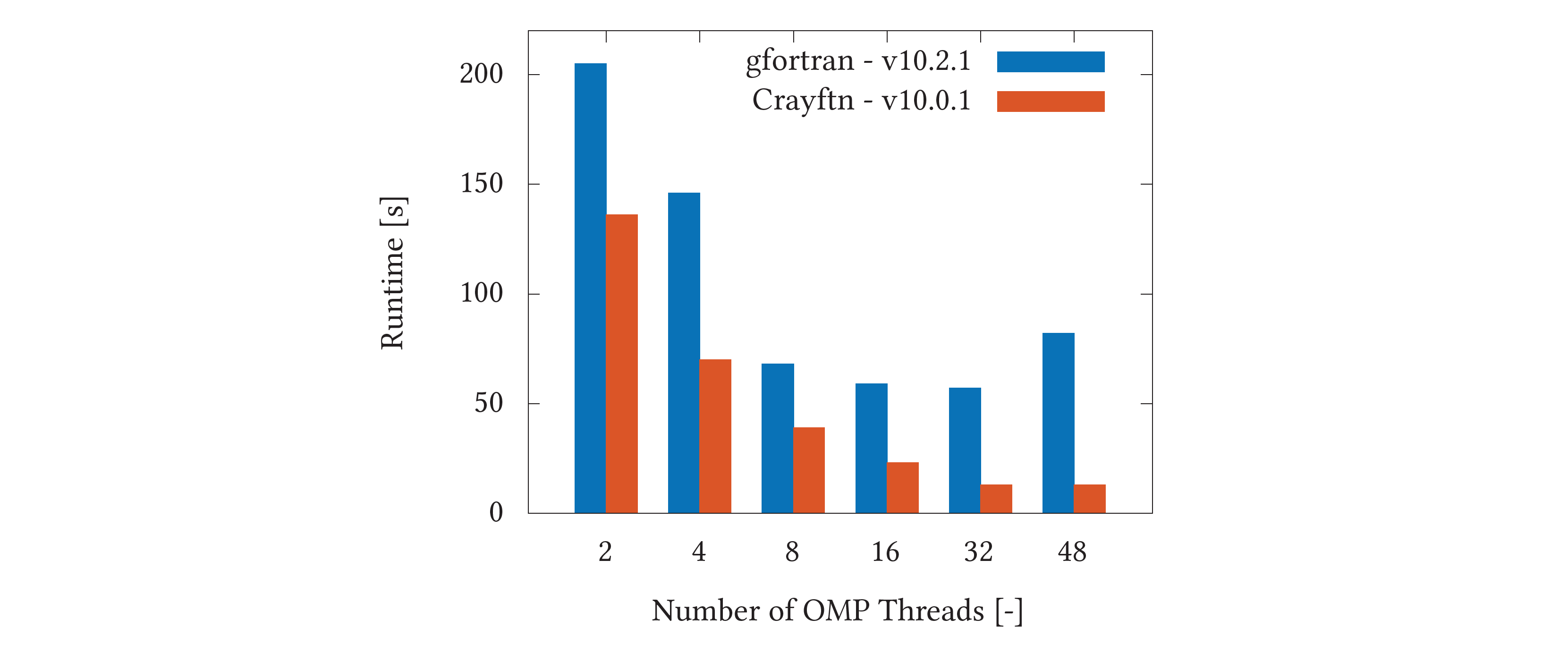}
    \caption{SWIM: Runtime results for the Cray compiler on A64FX and the GNU compiler on dual socket SkyLake x86.}
    \label{fig:SWIM-Skylake}
\end{figure}

SWIM ~\cite{Swim} is a weather forecasting benchmark (FORTRAN OpenMP) that solves the shallow-water equations using finite difference, and tracks FP speed, bandwidth, and cache characteristics, and is now included in the SPEC CPU2000 Benchmark suite.  In our experiment, we used the default test problem \verb+swim.ref.in+ as the sample problem on two different clusters of different architectures. One is the A64FX cluster “Ookami” and the other is a SkyLake x86 cluster from Stony Brook Exasca||ab. Seven different compilers were tested on Ookami. To avoid clutter and data overlap, we have chosen 3 representatives: ARM's GNU-based compiler based on GCC 9.3.0 , ARM's LLVM-based compiler based on 9.0.1, and Cray 10.0.1. The times are shown in Figure~\ref{fig:SWIM-Ookami}. Times on the X86 cluster using GNU compiler 10.2.1 and on Ookami using Cray 10.0.1 are shown in Fig~\ref{fig:SWIM-Skylake}. As the results shown in Fig~\ref{fig:SWIM-Ookami},  the Cray Compiler obtained the best performance on Ookami. Ookami has a clear performance advantage with a single socket of A64FX being 4.3x faster at 32 threads compared with dual socket SkyLake x86. We ascribe the slight slow down at 48 threads on SkyLake to oversubscription (there are only 40 physical cores). 

\subsubsection{PENNANT} PENNANT is an unstructured mesh, MPI+OpenMP, physics mini-application~\cite{LANLPennant} designed for advanced architecture research. 
It is a challenge the A64FX processor due to the lack of locality and the architectures 256~byte cache line. This is particularly evident with the largest of the mesh inputs, each having at least 3 billion points and sides, with at least 11 billion edges.
MPI handles
inter-node gather/scatters and reductions, while OpenMP is used for intra-node activities like mesh initialization, summing energy usage, and data allocation.
To explore the scalability of PENNANT on A64FX and compare its performance to other processors, we tested one of PENNANT's smaller tests (LeblancBig).

\begin{figure}[ht!]
    \centering
    \includegraphics[width=\columnwidth]{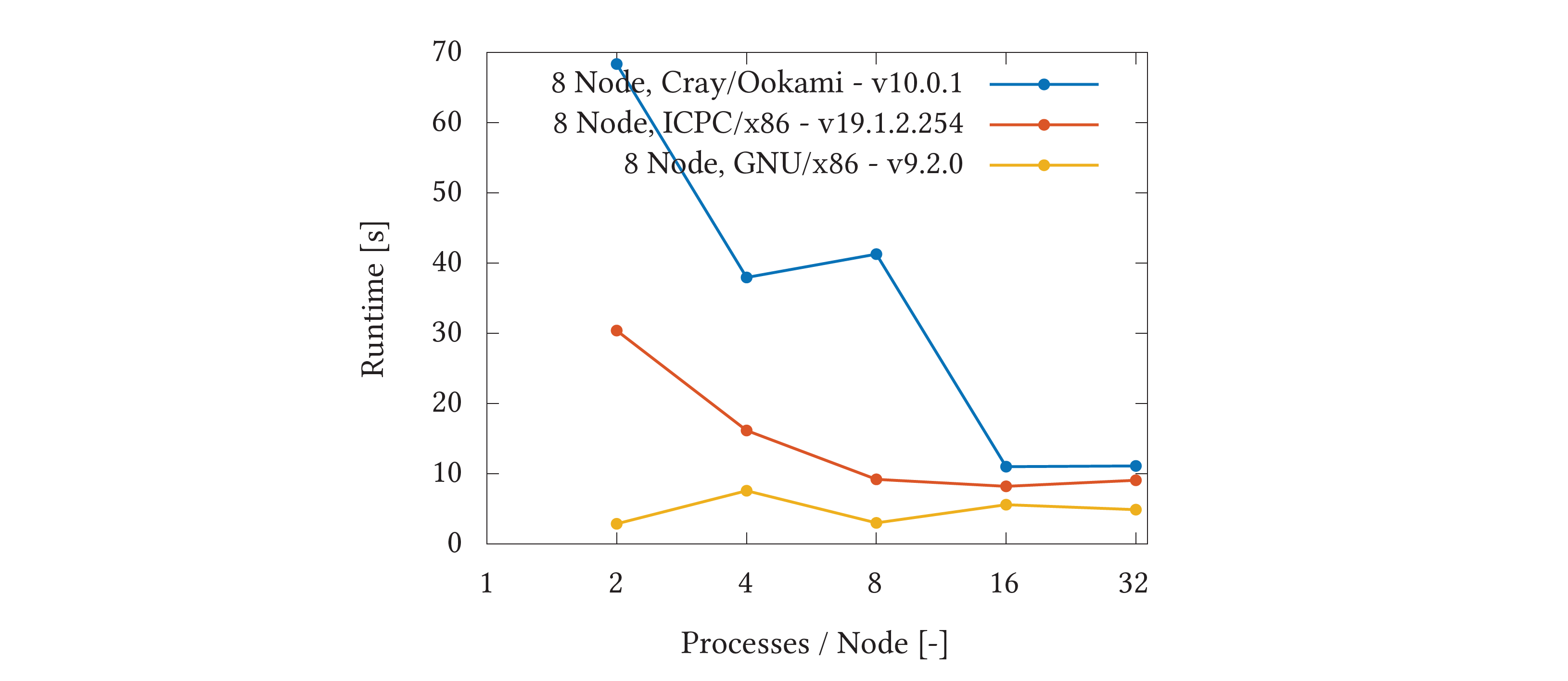}
    \caption{PENNANT: Run-time vs. no. cores between 3 different compilers with optimizations}
    \label{fig:PENNANT-MPI}
\end{figure}

MPI scaling experiments
were run on the Ookami and SBU SeaWulf clusters, using the Cray (Ookami), Intel, and GNU (SeaWulf) compiler toolchains. Tests used an 8 node configuration, running from 2 to 32 MPI ranks/node, and with one OpenMP thread per process. While run-times begin converging at larger process counts, 
the Intel system is faster, as shown in Figure ~\ref{fig:PENNANT-MPI}.
We also tested single-node, OpenMP scaling on Ookami, comparing it to an Intel Haswell processor, with experiments using thread counts from 1 to 64 and using the GNU compiler on both platforms to facilitate direct comparison. 
The runtime spike from 11 seconds to nearly 600 once more than 48 OpenMP threads are requested is attributed to oversubscription, which is handled less gracefully by A64FX due to the lack of hyperthreading and current settings for O/S thread scheduling policies .

\subsubsection{GROMACS}\cite{GROMACS} is a molecular dynamics simulation package for modeling biomolecular systems such as proteins, DNA, RNA, and lipid membrane. The code extensively utilizes features of the C++17 language. Due to this, we were only able to build binaries with GCC(10.2.0) but not with ARM(20.3) and Cray(10.0.1) compilers and with other versions of GCC failing due to language constructs.  SVE support was introduced to GROMACS in 2021 version by specialists from Research Organization for Information Science and Technology (Japan). Before that, GROMACS had optimization for ARM NEON. Below, we compare Ookami no-SIMD/NEON/SVE performances with no-SIMD/SSE2/AVX512 on dual socket 2 Intel Xeon Platinum 8160  (TACC Stampede-2).   From the GROMACS benchmark set\cite{GROMACSBench} we report results for a protein embedded into a membrane, consisting of 87k atoms. 

On a single Ookami node, the performance for no-SIMD, NEON and SVE versions are 3.6, 13.8 and 20.2 ns/day, respectively. On Stampede-2 system the performance for no-SIMD, SSE2 and AVX512 versions are 12.7, 47.8 and 75.6 ns/day. In both cases (Ookami and Stampede 2) the gain from utilizing wider SIMD is very similar, a fascinating, 3.8 times performance increase by moving from no-SIMD to 128-bit wide SIMD and  moderate 50\% increase by moving from 128-bits wide to 512-bits wide SIMD. What makes it interesting is that although Ookami and Stampede 2 theoretical flops are similar and aforementioned similar scaling on SIMD width increase, the absolute performance of Ookami is significantly smaller. It is not clear whether this is  due to architectural differences or immature compiler support (including the inability of GCC to vectorize math functions).

\subsubsection{FLASH} 
The FLASH code is a large, component-based package written primarily in modern FORTRAN that has a wide user base addressing a variety of multi-scale, multi-physics applications, especially astrophysics~\cite{Fryxetal00,calder.fryxell.ea:on,flash_development}.
The current release version of FLASH (4.6.2) is parallelized using MPI, and utilizes PARAMESH  \cite{macneice.olson.ea:paramesh}
to support its block-structured Adaptive Mesh Refinement (AMR) scheme.
FLASH is able to tune the amount of memory per processor it uses, and be paused and restarted on a different number of cores. All of these features make FLASH a promising platform to take advantage of the A64FX’s HBM, NUMA architecture, and SVE instructions.

We successfully tested multiple combinations of compilers and MPI implementations (MVAPICH 2.3.4 and OpenMPI 4.0.5), initially without optimization to assess ease of porting and correctness. The GCC (9.2.0 and 10.2.1) and NVIDIA (20.9) compilers worked out of the box, while using the ARM (20.3.0) and Cray (10.0.3) compilers just required setting proper options and MPI environment variables. We were also able to demonstrate strong scaling of our supernova simulation without optimization on 12--140 cores using the GCC 9.2.0 compiler with MVAPICH 2.3.4. However, the SeaWulf cluster currently provides over double the speedup.  Our next goal is to enable SVE, however the ARM and Cray compilers generate runtime errors, and the GCC compiler provides no significant speedup. The application chosen for this study was an explosive astrophysical event known as a thermonuclear or Type Ia supernova. We are exploring the dimmest of these events, known as a Type Iax, which is thought to be produced by a pure deflagration (subsonic burning front) occurring in a special ``hybrid" white dwarf  \cite{Foley_2013,Kromer_2015}, which leaves behind a bound remnant. 


\subsection{XDMoD}

\begin{figure}[h!]
    \centering
    \includegraphics[width=\columnwidth]{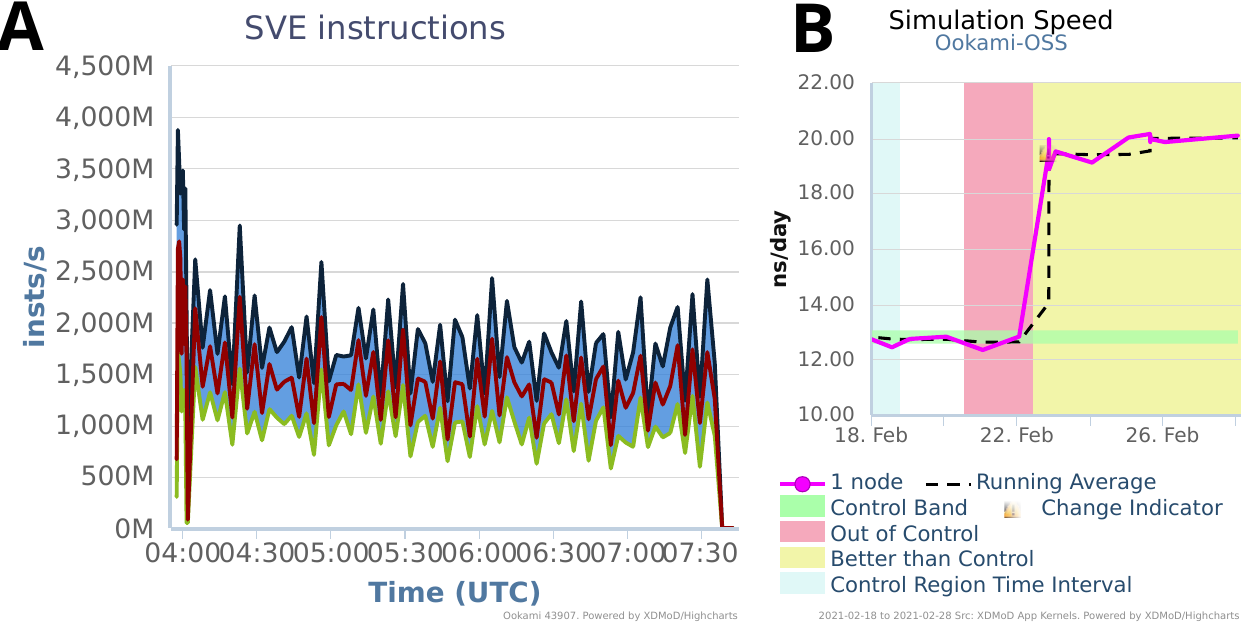}
    \caption{A) XDMoD job viewer plot of the SVE instruction rate for a 150 node run of the MADNESS software. At each time point the SVE instruction rate for the compute nodes with the maximum, median and minimum value is shown. B) Daily execution of GROMACS application kernel shows the performance improvement due to version update from 2020.4 to 2021 (first release with SVE support).}
    \label{fig:xdmod_job_viewer_sve}
\end{figure}

Ookami is monitored with XDMoD~\cite{Palmer:2015}, with the
XDMoD job performance and application kernel modules also installed.
The XDMoD instance is hosted in CCR's cloud computing cluster at the
University at Buffalo. Ookami users can access the XDMoD instance
using their Stony Brook accounts via Globus Auth.  The default XDMoD install worked without requiring
any customization. However, we did modify the XDMoD software to monitor
A64FX-specific metrics.
Performance data are collected using 
performance co-pilot (PCP)~\cite{pcp}. We used the PCP version
supplied with CentOS 8 but recompiled the performance monitoring
module against the latest version of \texttt{libpfm4} to 
log data from the A64FX hardware counters.
We also backported the latest version of the module that records
information from Lustre because Ookami uses a newer version of the
Lustre driver than is supplied with CentOS.
The XDMoD job summarization software was updated to handle the ARM 
SVE hardware counters, and time-series plots of the instruction rate
are available in XDMoD's job viewer. Figure~\ref{fig:xdmod_job_viewer_sve}.A shows
an example plot of the SVE instruction rate exported from the XDMoD
job viewer.

Within XDMoD, application kernels are used to proactively monitor HPC resource performance by daily benchmarks. For Ookami, a goal is to see how the performance of benchmarks and real applications change as the compiler
toolchains improve.  Figure~\ref{fig:xdmod_job_viewer_sve}.B illustrates a performance improvement for the GROMACS application. 
We plan to add a power data collector so that we can analyse job level
power usage in XDMoD, and also adding new job-efficiency analytics that take into
account the extra ARM hardware counters, e.g., to identify floating-point intensive codes not using SVE.

\section{Summary}

In conclusion, we have had very positive initial experience with A64FX.  It should be viewed as a ``leadership processor'' that trades high performance and high power efficiency on a large class of well-vectorized scientific applications for reduced performance (especially if not vectorized) and reduced applicability (primarily due to memory capacity) on more general codes.  The peak processor vector speed and peak memory bandwidth are indeed readily accessible to compiled codes that are well vectorized and pay attention to localizing memory references within a CMG.  The latter is readily accomplished by running four multi-threaded MPI processes per node, with one per CMG. So far, it is living up to the expectation that most such software can deliver great performance out of the box.  However, the relative immaturity of the SVE software ecosystem (again noting that we do not yet have the Fujitsu stack) makes it hard to generalize this statement. 

\begin{acks}

Ookami is supported by the National Science Foundation (NSF) grant OAC 1927880, XDMoD by NSF OAC 1445806, and Seawulf by NSF OAC 1531492. OpenSHMEM  is funded through Los Alamos National Laboratory, Grant \#367958.
FLASH was developed in part by the US Department of Energy (DOE) NNSA-ASC and OSC-ASCR-supported Flash Center for Computational Science at the University of Chicago. Work involving supernovae research was supported in part by the DOE under grant DE-FG02-87ER40317. We would like to thank Total Exploration and Production Research and Technologies for their support of experimentation using MiniMod. 
Use of the Oak Ridge Leadership Computing Facility at the Oak Ridge National Laboratory was supported by the Office of Science of the DOE under Contract No. DE-AC05-00OR22725. Use of XSEDE systems is supported by XSEDE grant TG-CCR120014.
We also thank the MVAPICH and Open-MPI teams for their assistance in tuning and deploying their software.

\end{acks}

\balance
\bibliographystyle{ACM-Reference-Format}
\bibliography{bibliography.bib}
\end{document}